\documentclass[12pt]{article}
\usepackage{epstopdf}
\usepackage{amsmath,amssymb,amsfonts,graphicx}

\begin{document}

\thispagestyle{empty}
\renewcommand{\refname}{References}

\title{\bf Casimir effect with quantized charged spinor matter in \\
background magnetic field}


\author{Yu. A. Sitenko$^{1,2}$}

\date{}

\maketitle
\begin{center}
$^{1}$ Bogolyubov Institute for Theoretical Physics,\\
National Academy of Sciences of Ukraine,\\
14-b Metrologichna Street, 03680 Kyiv, Ukraine\\
$^{2}$ Institute for Theoretical Physics, University of Bern,\\
Sidlerstrasse 5, CH-3012 Bern, Switzerland

\end{center}

\begin{abstract}
We study the influence of a background uniform magnetic field and
boundary conditions on the vacuum of a quantized charged spinor
matter field confined between two parallel neutral plates; the
magnetic field is directed orthogonally to the plates. The
admissible set of boundary conditions at the plates is determined
by the requirement that the Dirac Hamiltonian operator be
self-adjoint. It is shown that, in the case of a sufficiently
strong magnetic field and a sufficiently large separation of the
plates, the generalized Casimir force is repulsive, being
independent of the choice of a boundary condition, as well as of
the distance between the plates. The detection of this effect
seems to be feasible in the foreseeable future.
\end{abstract}


\bigskip

\begin{center}
\end{center}

\bigskip
\medskip

\section{Introduction}

Zero-point oscillations in the vacuum of quantized matter fields
that are subject to boundary conditions have been studied
intensively over more than six decades since H.B.G.Casimir
\cite{Cas1,Cas2} predicted a force between grounded metal plates,
see reviews in \cite{Most,Mil,Bor}. The existence of this force is
one of the few macroscopic manifestations of quantum theory,
together with other remarkable phenomena such as superfluidity,
superconductivity, kaon and neutrino oscillations, spectrum of
black-body radiation. The Casimir force between material
boundaries has now been measured quite accurately, it agrees with
theoretical predictions, see, e.g., \cite{Dec,Lam}, as well as
other publications cited in \cite{Bor}, and this opens a way for
various applications in modern nanotechnology.

The Casimir force is closely related to the van der Waals force
between material bodies at such separation distances ($> 10^{-8}
\rm m$) that the retardation owing to the finiteness of the velocity
of light becomes important. In view of this, it seems that the
following two circumstances have to be clearly noted. The first
one is that, as long as the intermolecular van der Waals forces
are attractive, almost all experimental measurements reveal
the attractive Casimir force; an evidence for the repulsive
Casimir force has appeared just several years ago \cite{Mun}. The
second one is that, as long as the intermolecular van der Waals
forces are due to electromagnetic fluctuations, the Casimir effect
is caused by zero-point oscillations in the vacuum of the
quantized electromagnetic field. The Casimir effect with other
(nonelectromagnetic) quantized fields is mostly regarded as 
merely an academic exercise that could hardly be validated in
laboratory. However, the nonelectromagnetic fields can be
charged, and this opens a new prospect allowing one to consider
the Casimir effect as that caused by zero-point oscillations in
the vacuum of quantized charged matter fields in the presence of
material boundaries and a background (classical) electromagnetic
field inside the quantization volume. Whether the Casimir effect
of this kind is attractive or repulsive  --  we shall get an
answer in the present paper.

Let us start by recalling that the effect of the background
uniform electromagnetic field alone on the vacuum of quantized
charged matter was studied long ago, see
\cite{Hei1,Eul,Hei2,Wei,Schw} and review in \cite{Dun}. The case
of a background field filling the whole (infinite) space is hard
to be regarded as realistic, whereas the case of a background
field confined to the bounded quantization volume for charged
matter looks much more plausible, it can even be regarded as
realizable in laboratory. Moreover, there is no way to detect the
energy density which is induced in the vacuum in the first case,
whereas the pressure from the vacuum onto the boundary, resulting
in the second case, is in principle detectable. One may suggest
intuitively that the pressure, at least in certain circumstances,
is positive, i.e. directed from the inside to the outside of the
quantization volume. A natural question is then, whether the
pressure depends on a boundary condition imposed on the quantized
charged matter field at the boundary?

Thus, an issue of a choice of boundary conditions acquires a
primary importance, requiring a thorough examination. It should be
recalled that, in the conventional case of the Casimir effect with
the quantized electromagnetic field, there exist physical
motivations for different boundary conditions, for instance,
corresponding to metallic or dielectric plates, see, e.g.,
\cite{Bor}. Such motivations seem to be lacking for the case of
quantized charged matter fields, but it was not distressing as
long as this case, as we have already mentioned, was regarded as a
purely academic one. Otherwise, in a situation which is supposed
to be physically sensible, one should be guided by general
principles, such as comprehensiveness and mathematical
consistency, while seeking out boundary conditions. Namely, 
in the context of first-quantized theory, a
quest is for the operator of a physical observable to be
self-adjoint rather than Hermitian. This is stipulated by the mere
fact that a multiple action is well defined for the self-adjoint
operator only, allowing for the construction of functions of the
operator, such as evolution, resolvent, zeta-function and heat
kernel operators, with further implications upon second
quantization. Whether the quest can be fulfilled successfully is
in general determined by the Weyl -- von Neumann theory of
self-adjoint operators, see, e.g., \cite{Neu,Akhi}. Thus, the
requirement of the self-adjointness for the operator of
one-particle energy (Dirac Hamiltonian operator in the case of
quantized relativistic spinor fields) renders the most general set
of boundary conditions, which may be further restricted by
additional physical considerations.

To avoid a misunderstanding, let us emphasize once more that quantized
matter fields are assumed to be confined within the boundaries,
and an issue of what is out of the boundaries is not touched upon.
In a sense, this setup is the same as that in modeling hadrons as
bags containing the quark matter, see \cite{Joh,Has}. In
distinction to the conventional setup for the Casimir effect, the
impact of background fields on confined quantized matter fields is
added along the lines discussed above. This generalization implies
that the boundaries perceive an additional physical meaning,
serving as a source of background fields which are inside
the quantization volume.

In the present paper, we consider the Casimir effect in the
generalized setup for a quantized charged spinor matter
field in the background of an external uniform magnetic field;
both the quantized and external fields are confined between two
parallel plates, and the external field is orthogonal to the
plates. It should be noted that a similar problem has been studied
more than a decade ago \cite{Cou,Eli,Ost} in a setup which is
somewhat closer to the conventional setup for the Casimir effect.
Namely, the authors of \cite{Cou,Eli,Ost} assume that both the
quantized and external fields are not confined between the plates
but extend outside; the plates in their setup are regarded as the
places where constraints on quantized fields are imposed, rather
than as the real boundaries of the quantization volume. One of the
purposes of the present paper is to compare the results obtained
in these two different physical situations. According to
\cite{Cou,Eli,Ost}, there is no room for the validation of the
aforementioned intuitive suggestion: the pressure in all
circumstances is negative, i.e. the plates are attracted. On the
contrary, by studying a response of the vacuum of the confined
quantized spinor matter field on the external magnetic field with
the strength lines terminating at the plates, I shall show that,
in the case of a sufficiently strong magnetic field and a
sufficiently large separation of the plates, the pressure from the
vacuum onto the plates is positive, being independent of the
choice of a boundary condition and even of the distance between
the plates.

In the next section we consider in general the problem of the
self-adjointness for the Dirac Hamiltonian operator. In Section 3
we discuss the vacuum energy which is induced by an external
uniform magnetic field and compare the appropriate expressions for
the cases of the unbounded quantization volume and the
quantization volume bounded by two parallel plates. A condition
determining the spectrum of wave number vector in the direction of
the magnetic field is chosen in Section 4. Expressions for the
Casimir energy and force are obtained in Section 5. The
conclusions are drawn and discussed in Section 6. Some details of
the derivation of results are given in Appendices A and B.

\section{Self-adjointness of the Dirac Hamiltonian operator}

Defining a scalar product as
$$
(\xi,\chi)=\int\limits_{\Omega}{\rm d}^3r\,{\xi}^{\dag}\chi\, ,
$$
we get, using integration by parts,
$$
(\xi,H\chi)=(H^{\dag}\xi,\chi)- {\rm
i}\int\limits_{\partial{\Omega}}{\rm
d}\boldsymbol{\sigma}\cdot\bar{\xi}\boldsymbol{\gamma}\chi,\eqno(1)
$$
where $\bar{\xi}={\xi}^{\dag}\gamma^0$ and
$$
H=H^{\dag}=-{\rm
i}\gamma^0\boldsymbol{\gamma}\cdot(\boldsymbol{\partial}-{\rm
i}e\mathbf{A})+eA_{0}+\gamma^0 m,\eqno(2)
$$
is the formal expression for the Dirac Hamiltonian operator in an
external electromagnetic field (natural units ${\hbar}=c=1$ are
used), $\partial{\Omega}$ is a two-dimensional surface bounding
the three-dimensional spatial region $\Omega$. Operator $H$ is
Hermitian (or symmetric in mathematical parlance),
$$
(\xi,H\chi)=(H^{\dag}\xi,\chi),\eqno(3)
$$
if
$$
\int\limits_{\partial{\Omega}}{\rm
d}\boldsymbol{\sigma}\cdot\bar{\xi}\boldsymbol{\gamma}\chi = 0.
\eqno(4)
$$
It is almost evident that the latter condition can be satisfied by
imposing different boundary conditions for $\chi$ and $\xi$. But,
a nontrivial task is to find a possibility that a boundary
condition for $\xi$ is the same as that for $\chi$; then the
domain of definition of $H^{\dag}$ (set of functions $\xi$)
coincides with that of $H$ (set of functions $\chi$), and operator
$H$ is called self-adjoint. The action of a self-adjoint operator
results in functions belonging to its domain of definition only,
and, therefore, a multiple action and functions of such an
operator can be consistently defined.

Condition (4) is certainly fulfilled when the integrand in (4)
vanishes, i.e.
$$
\boldsymbol{n}\cdot\bar{\xi}\boldsymbol{\gamma}\chi|_{\mathbf{r}
\in
\partial{\Omega}}=0,\eqno(5)
$$
where $\boldsymbol{n}$ is the unit normal which may be chosen as
pointing outward to the boundary. To fulfill the latter condition, we impose
the same boundary condition for $\chi$ and $\xi$ in the form
$$
\chi|_{\mathbf{r}\in \partial{\Omega}}=K\chi|_{\mathbf{r}\in
\partial{\Omega}},\quad \xi|_{\mathbf{r}\in
\partial{\Omega}}=K\xi|_{\mathbf{r}\in \partial{\Omega}},\eqno(6)
$$
where $K$ is a matrix (element of the Clifford algebra) which is
determined by two conditions:
$$
K^{2}=I\eqno(7)
$$
and
$$
K^{\dag}\gamma^0\boldsymbol{n}\cdot\boldsymbol{\gamma}K=-\gamma^0\boldsymbol{n}\cdot\boldsymbol{\gamma}.\eqno(8)
$$
Using the standard representation for $\gamma$-matrices,
$$
\gamma^0=\begin{pmatrix}
I&0\\
0&-I
\end{pmatrix},\qquad
\boldsymbol{\gamma}=\begin{pmatrix}
0&\boldsymbol{\sigma}\\
-\boldsymbol{\sigma}&0
\end{pmatrix}\eqno(9)$$
($\sigma^{1},\sigma^{2}$ and $\sigma^{3}$ are the Pauli matrices),
one can get
$$
K=\begin{pmatrix}
0&{\varrho}^{-1}\\
\varrho&0
\end{pmatrix},\eqno(10)
$$
where condition
$$
\boldsymbol{n}\cdot\boldsymbol{\sigma}
\varrho=-{\varrho}^{\dag}\boldsymbol{n}\cdot\boldsymbol{\sigma}
\eqno(11)
$$
defines $\varrho$ as a rank-2 matrix depending on four arbitrary
parameters \cite{Wie}. An explicit form for matrix $K$ is
$$
K=\frac{(1+u^2-v^2-{\boldsymbol{t}}^2)I+(1-u^2+v^2+{\boldsymbol{t}}^2)\gamma^0}{2{\rm
i}(u^2-v^2-{\boldsymbol{t}}^2)}(u\boldsymbol{n}\cdot\boldsymbol{\gamma}+v\gamma^{5}-{\rm
i}\boldsymbol{t}\cdot\boldsymbol{\gamma}),\eqno(12)
$$
where $\gamma^{5}=-{\rm i}\gamma^0\gamma^1\gamma^2\gamma^3$, and
$\boldsymbol{t}=(t^1,t^2)$ is a two-dimensional vector which is
tangential to the boundary. Hence, the boundary condition ensuring
the self-adjointness of operator $H$ (2) is written explicitly as
$$
\biggl\{I-\frac{I(\cosh^2\tilde{\vartheta}+1)-\gamma^0\sinh^2\tilde{\vartheta}}{2{\rm
i}\cosh\tilde{\vartheta}}[\boldsymbol{n}\cdot
\boldsymbol{\gamma}\cosh\vartheta+\gamma^{5}\sinh\vartheta\cos\theta
$$
$$
-{\rm
i}(\gamma^{1}\cos\phi+\gamma^{2}\sin\phi)\sinh\vartheta\sin\theta]\biggr\}\chi|_{\mathbf{r}\in
\partial{\Omega}}=0 \eqno(13)
$$
(the same condition is for $\xi$), where
$$
[\boldsymbol{n}\cdot\boldsymbol{\gamma},\,\gamma^1]_{+}=
[\boldsymbol{n}\cdot\boldsymbol{\gamma},\,\gamma^2]_{+}=[\gamma^1,\gamma^2]_{+}=0,
\eqno(14)
$$
and we have employed parametrization
$$
u=\cosh\tilde{\vartheta}\cosh\vartheta, \quad
v=\cosh\tilde{\vartheta}\sinh\vartheta\cos\theta,
$$
$$
t^1=\cosh\tilde{\vartheta}\sinh\vartheta\sin\theta\cos\phi, \quad
t^2=\cosh\tilde{\vartheta}\sinh\vartheta\sin\theta\sin\phi,
$$
$$
-\infty<\vartheta<\infty, \quad 0\leq\tilde{\vartheta}<\infty,
\quad 0\leq\theta<\pi, \quad 0\leq\phi<2\pi.\eqno(15)
$$
Parameters of the boundary condition, $\vartheta$,
$\tilde{\vartheta}$, $\theta$ and $\phi$, can be interpreted as
the self-adjoint extension parameters. It should be noted that, in
addition to (5), the following combination of $\chi$ and $\xi$ is
also vanishing at the boundary:
$$
\frac{1}{2}\bar{\xi}[I(\cosh^2\tilde{\vartheta}+1)+\gamma^0\sinh^2\tilde{\vartheta}][I\cosh\vartheta-\boldsymbol{n}\cdot
\boldsymbol{\gamma}\gamma^{5}\sinh\vartheta\cos\theta
$$
$$
+{\rm i}\boldsymbol{n}\cdot
\boldsymbol{\gamma}(\gamma^{1}\cos\phi+\gamma^{2}\sin\phi)\sinh\vartheta\sin\theta]\chi|_{\mathbf{r}\in
\partial{\Omega}}=0.\eqno(16)
$$

Clearly, parametrization (15) is relevant for the case of $1\leq
u^2-v^2-{\boldsymbol{t}}^2<\infty$ only. The case of $0<
u^2-v^2-{\boldsymbol{t}}^2 \leq 1$ corresponds to the imaginary
values of $\tilde{\vartheta}$: ${\rm Re}\tilde{\vartheta}=0$,
$0\leq{\rm Im}\tilde{\vartheta}<\pi/2$. At
$\vartheta=\tilde{\vartheta}=\theta=\phi=0$ one obtains the
well-known MIT bag boundary condition \cite{Cho1,Cho2}, see
reviews in \cite{Joh,Has}:
$$
(I+{\rm
i}\boldsymbol{n}\cdot\boldsymbol{\gamma})\chi|_{\mathbf{r}\in
\partial{\Omega}}=(I+{\rm
i}\boldsymbol{n}\cdot\boldsymbol{\gamma})\xi|_{\mathbf{r}\in
\partial{\Omega}}=0,\eqno(17)
$$
and relation (16) takes form
$$
\bar{\xi}\chi|_{\mathbf{r}\in
\partial{\Omega}}=0.\eqno(18)
$$
The case of $-\infty < u^2-v^2-{\boldsymbol{t}}^2 < 0$ is hard to
be regarded as physically acceptable, since a link to the MIT bag
boundary condition is lacking.

It should be noted that, in the case of the two-dimensional
Dirac-Weyl hamiltonian operator emerging in the framework of the
tight-binding model description of long-wavelength electronic
excitations in graphene, the most general boundary condition is
also four-parametric \cite{Bee}, but the $K$-matrix is chosen to be Hermitian
in this case.

Returning to the case of operator $H$ (2), we note that, if the boundary is disconnected, consisting of several connected
components, $\partial{\Omega} = \bigcup\limits_{J}
\,\,\partial{\Omega}^{(J)}$, then there are four ($\vartheta_{J}$,
$\tilde{\vartheta}_{J}$, $\theta_{J}$ and $\phi_{J}$) self-adjoint
extension parameters  corresponding to each of the components,
$\partial{\Omega}^{(J)}$. However, if some symmetry is present,
then the number of self-adjoint extension parameters can be
diminished. For instance, let us consider spatial region $\Omega$
which is bounded by two noncompact noncontiguous surfaces,
$\partial{\Omega}^{(+)}$ and $\partial{\Omega}^{(-)}$. Choosing
coordinates $\mathbf{r}=(x,y,z)$ in such a way that $x$ and $y$
are tangential to the boundary, while $z$ is normal to it, we
identify the position of $\partial{\Omega}^{(\pm)}$ with, say,
$z=\pm{a/2}$. If region $\Omega$ is invariant under rotations
around a normal to the boundary surfaces (that is the case of a
region bounded by parallel planes), then the boundary condition
should be independent of the components of the
$\boldsymbol{\gamma}$-vector, which are tangential to the boundary,
i.e.
$$
\theta_{+}=\theta_{-}=0. \eqno(19)
$$
Operator $H$ (2) acting on functions which are defined in such a
region is self-adjoint if condition
$$
[I-\frac{I(\cosh^2\tilde{\vartheta}_{\pm}+1)-\gamma^0\sinh^2\tilde{\vartheta}_{\pm}}{2{\rm
i}\cosh\tilde{\vartheta}_{\pm}}({\pm}{\gamma}^3\cosh\vartheta_{\pm}+\gamma^{5}\sinh\vartheta_{\pm})]\chi|_{z=\pm{a/2}}=0,
\eqno(20)
$$
holds (with the same condition holding for $\xi$). The latter
ensures the fulfilment of constraint
$$
\bar{\xi}\gamma^{3}\chi|_{z=\pm{a/2}}=0,\eqno(21)
$$
as well as of relation
$$
\frac{1}{2}\bar{\xi}[I(\cosh^2\tilde{\vartheta}_{\pm}+1)+\gamma^0\sinh^2\tilde{\vartheta}_{\pm}]
(I\cosh\vartheta_{\pm} \mp
{\gamma}^3\gamma^{5}\sinh\vartheta_{\pm})\chi|_{z=\pm{a/2}}=0.
\eqno(22)
$$

    \section{Induced vacuum energy in the magnetic field
    background}

The operator of a spinor field which is quantized in a static
background is presented in the form
$$\hat{\Psi}(t,\mathbf{r})=\sum\!\!\!\!\!\!\!\!\!\!\!\int\limits_{E_{\lambda}>0}{\rm e}^{-{\rm i}E_{\lambda}t}\psi_{\lambda}(\mathbf{r})\hat{a}_{\lambda}
+\sum\!\!\!\!\!\!\!\!\!\!\!\int\limits_{E_{\lambda}<0}{\rm
e}^{-{\rm
i}E_{\lambda}t}\psi_{\lambda}(\mathbf{r})\hat{b}^{\dag}_{\lambda},\eqno(23)
$$
where
$\hat{a}^{\dag}_{\lambda}$ and $\hat{a}_{\lambda}$
($\hat{b}^{\dag}_{\lambda}$ and $\hat{b}_{\lambda}$) are the
spinor particle (antiparticle) creation and destruction operators,
satisfying anticommutation relations
$$
[\hat{a}_\lambda,\hat{a}_{\lambda'}^\dagger]_+=[\hat{b}_\lambda,\hat{b}_{\lambda'}^\dagger]_+=\left\langle
\lambda|\lambda'\right\rangle,\eqno(24)
$$
wave functions $\psi_{\lambda}(\textbf{r})$ form a
complete set of solutions to the stationary Dirac equation
$$
H\psi_{\lambda}(\mathbf{r})=E_{\lambda}\psi_{\lambda}(\mathbf{r});\eqno(25)
$$
$\lambda$ is the set of parameters (quantum numbers) specifying a
one-particle state with energy $E_{\lambda}$; symbol
$\sum\!\!\!\!\!\!\!\int\,$ denotes summation over discrete and
integration (with a certain measure) over continuous values of
$\lambda$. Ground state $|\texttt{vac}>$ is defined by condition
$$
\hat{a}_\lambda|\texttt{vac}>=\hat{b}_\lambda|\texttt{vac}>=0.\eqno(26)
$$
The temporal component of the operator of the energy-momentum tensor
is given by expression
$$
\hat{T}^{00}=\frac{\rm{i}}{4}[\hat{\Psi}^{\dag}({\partial_0}\hat{\Psi})-
({\partial_0}\hat{\Psi}^{T})\hat{\Psi}^{{\dag}T}-({\partial_0}\hat{\Psi}^{\dag})\hat{\Psi}+
\hat{\Psi}^{T}({\partial_0}\hat{\Psi}^{{\dag}T})],\eqno(27)
$$where superscript $T$ denotes a transposed spinor. Consequently, the formal expression for the vacuum expectation
value of the energy
density is
$$
\varepsilon=<\texttt{vac}|\hat{T}^{00}|\texttt{vac}>=
-\frac{1}{2}\sum\!\!\!\!\!\!\!\!\!\int
\,|E_{\lambda}|\psi_{\lambda}^{\dag}(\textbf{r})\psi_{\lambda}(\textbf{r}).\eqno(28)
$$

 Let us consider the quantized charged massive spinor field
in the background of a static uniform magnetic field; then $A_{0}=0$
and the gauge in $H$ (2) can be chosen as $\mathbf{A}=(-yB,0,0)$,
where $B$ is the magnetic field strength which is directed along the
$z$-axis in Cartesian coordinates $\mathbf{r}=(x,y,z)$. The
one-particle energy spectrum is
$$
E_{nk}=\pm\omega_{nk},\eqno(29)
$$where
$$
\omega_{nk}=\sqrt{2n|eB|+k^{2}+m^{2}},\;-\infty<k<\infty,\;n=0,1,2,...\,
,\eqno(30)
$$
$k$ is the value of the wave number vector along the $z$-axis, and
$n$ numerates the Landau levels. Although a solution to the Dirac
equation in the background of a static uniform magnetic field is
well-described in the literature, see, e.g., \cite{Akhie}, we list
it below for self-consistency. Taking $eB>0$ for definiteness, the
solution with positive energy, $E_{nk}=\omega_{nk}$, is
$$\psi_{qnk}(\mathbf{r})=\frac{{\rm e}^{{\rm i}qx}{\rm e}^{{\rm i}kz}}{2\pi\sqrt{2\omega_{nk}(\omega_{nk}+m)}}\left[C_{1}\begin{pmatrix} (\omega_{nk}+m)Y_{n}^{q}(y) \\ 0 \\ kY_{n}^{q}(y) \\ \sqrt{2neB} Y_{n-1}^{q}(y) \end{pmatrix}\right.$$
$$\left.+C_{2}\begin{pmatrix} 0 \\ (\omega_{nk}+m)Y_{n-1}^{q}(y) \\
\sqrt{2neB} Y_{n}^{q}(y) \\ -kY_{n-1}^{q}(y)
\end{pmatrix}\right],\quad n\geq1 \eqno(31)
$$
and
$$\psi^{(0)}_{q0k}(\mathbf{r})=\frac{{\rm e}^{{\rm i}qx}{\rm e}^{{\rm i}kz}}{2\pi\sqrt{2\omega_{0k}(\omega_{0k}+m)}}
C_{0}Y_{0}^{q}(y)\begin{pmatrix} \omega_{0k}+m \\ 0 \\ k \\ 0
\end{pmatrix},\eqno(32)
$$
where $-\infty<q<\infty$ and
$$Y_{n}^{q}(y)=\sqrt{\frac{(eB)^{1/2}}{2^{n}n!\pi^{1/2}}}\exp{\left[-\frac{eB}{2}\left(y+\frac{q}{eB}\right)^{2}\right]
H_{n}\left[\sqrt{eB}\left(y+\frac{q}{eB}\right)\right]},\eqno(33)
$$
$H_n(u)$ is the Hermite polynomial. The solution with
negative energy, $E_{nk}=-\omega_{nk}$, is given in Appendix A,
see (A.1) and (A.2). The case of $eB<0$ is obtained by charge
conjugation, i.e. changing $eB\rightarrow{-eB}$ and multiplying
the complex conjugates  of the previous expressions by ${\rm i}
{\gamma}^2$ (the energy sign is changed to the opposite).

In the case of $n\geq1$, two linearly independent solutions,
$\psi_{qnk}^{(1)}(\mathbf{r})$ and $\psi_{qnk}^{(2)}(\mathbf{r})$,
are orthogonal, if the appropriate coefficients, $C_{j}^{(1)}$ and
$C_{j}^{(2)} \, (j=1,2)$, obey condition
$$
\sum\limits_{j=1,2}C_{j}^{(1)*}C_{j}^{(2)}=0.\eqno(34)
$$
By imposing further condition
$$
\sum\limits_{j=1,2}|C_{j}^{(j')}|^{2}=|C_{0}|^{2}=1,\quad
j'=1,2,\eqno(35)
$$
we arrive at the wave functions satisfying the requirements of
orthonormality
$$\int{\rm d}^{3}r\,\psi^{(j)\dag}_{qnk}(\mathbf{r})\psi_{q'n'k'}^{(j')}(\mathbf{r})=
\delta_{jj'}\delta_{nn'}\delta(q-q')\delta(k-k'), \quad
j,j'=0,1,2\eqno(36)
$$
and completeness
$$\sum\limits_{{\rm sgn}(E_{nk})}\,\int\limits_{-\infty}^{\infty}{\rm{d}}q\int\limits_{-\infty}^{\infty}{\rm{d}}k
\left[\psi^{(0)}_{q0k}(\mathbf{r})\psi^{(0)\dag}_{q0k}(\mathbf{r'})+
\sum\limits_{n=1}^{\infty}\sum\limits_{j=1,2}\psi^{(j)}_{qnk}(\mathbf{r})\psi^{(j)\dag}_{qnk}(\mathbf{r'})\right]=
I\delta(\mathbf{r}-\mathbf{r'}).\eqno(37)
$$
With the use of relation
$$\int\limits_{-\infty}^{\infty}{\rm{d}}q\,\left[Y^{q}_{n}(y)\right]^2=|eB|,\eqno(38)
$$the formal expression for the vacuum expectation value of the energy density in the uniform magnetic field is readily obtained:
$$\varepsilon^{\infty}=-\frac{|eB|}{2\pi^{2}}\int\limits_{-\infty}^{\infty}{\rm d}k
\sum\limits_{n=0}^{\infty}i_{n}\omega_{nk},\eqno(39)
$$
where $i_{n}=1-\frac{1}{2} \delta_{n0}$; the superscript on the
left-hand side indicates that the magnetic field fills the whole
(infinite) space. The integral and the sum in (39) are divergent
and require regularization and renormalization. This problem has
been solved long ago by Heisenberg and Euler \cite{Hei2} (see also
\cite{Schw}), and we just list here their result
$$\varepsilon^{\infty}_{\rm ren}=\frac{1}{8\pi^{2}}\int\limits_{0}^{\infty}\frac{{\rm d}\tau}{\tau}{\rm e}^{-\tau}
\left[\frac{eBm^{2}}{\tau}\coth\left(\frac{eB\tau}{m^{2}}\right)-\frac{m^{4}}{\tau^2}
-\frac{1}{3}e^{2}B^{2}\right];\eqno(40)
$$note that the renormalization procedure involves subtraction at
$B=0$ and renormalization of the charge.

Let us turn now to the quantized charged massive spinor field in
the background of a static uniform magnetic field in spatial
region $\Omega$ bounded by two parallel surfaces
$\partial{\Omega}^{(+)}$ and $\partial{\Omega}^{(-)}$; the
position of $\partial{\Omega}^{(\pm)}$ is identified with
$z=\pm{a/2}$, and the magnetic field is orthogonal to the
boundary. In addition to the plane wave propagating with wave
number vector $k$ along the $z$-axis, see (31) and (32), let us
consider also the plane wave propagating in the opposite
direction, which in the case of $eB>0$ takes form
$$\psi_{qn-k}(\mathbf{r})=\frac{{\rm e}^{{\rm i}qx}{\rm e}^{-{\rm i}kz}}{2\pi\sqrt{2\omega_{nk}(\omega_{nk}+m)}}\left[\tilde{C}_{1}\begin{pmatrix} (\omega_{nk}+m)Y_{n}^{q}(y) \\ 0 \\ -kY_{n}^{q}(y) \\ \sqrt{2neB}Y_{n-1}^{q}(y) \end{pmatrix}\right.$$
$$\left.+\tilde{C}_{2}\begin{pmatrix} 0 \\ (\omega_{nk}+m)Y_{n-1}^{q}(y) \\
\sqrt{2neB}Y_{n}^{q}(y) \\ kY_{n-1}^{q}(y)
\end{pmatrix}\right], \quad n\geq1\eqno(41)
$$
and
$$\psi^{(0)}_{q0-k}(\mathbf{r})=\frac{{\rm e}^{{\rm i}qx}{\rm e}^{-{\rm i}kz}}{2\pi\sqrt{2\omega_{0k}(\omega_{0k}+m)}}
\tilde{C}_{0}Y_{0}^{q}(y)\begin{pmatrix} \omega_{0k}+m \\ 0 \\ -k
\\ 0 \end{pmatrix}.\eqno(42)
$$
Then the solution to (25) in region $\Omega$ is chosen as a
superposition of two plane waves propagating in opposite
directions,
$$
\psi_{qnl}(\mathbf{r})=\psi_{qnk_{l}}(\mathbf{r})+\psi_{qn-k_{l}}(\mathbf{r}),\eqno(43)
$$
where all restrictions on the values of coefficients $C_{j}$ and
$\tilde{C}_{j} \; (j=0,1,2)$ are withdrawn for a while, and the
values of wave number vector $k_{l} \; (l=0,\pm1,\pm2, ... )$ are
determined from boundary condition, see (20),
$$
[I-\frac{I(\cosh^2\tilde{\vartheta}_{\pm}+1)-\gamma^0\sinh^2\tilde{\vartheta}_{\pm}}{2{\rm
i}\cosh\tilde{\vartheta}_{\pm}}({\pm}{\gamma}^3\cosh\vartheta_{\pm}+\gamma^{5}\sinh\vartheta_{\pm})]
\psi_{qnl}(\mathbf{r})|_{z=\pm{a/2}}=0.
 \eqno(44)
$$

The last condition can be written as a set of conditions on the
coefficients:
$$
\left\{\begin{array}{l}M_{11}^{(n)}C_{1}+M_{12}^{(n)}C_{2}+M_{13}^{(n)}\tilde{C}_{1}+M_{14}^{(n)}\tilde{C}_{2}=0
\\
M_{21}^{(n)}C_{1}+M_{22}^{(n)}C_{2}+M_{23}^{(n)}\tilde{C}_{1}+M_{24}^{(n)}\tilde{C}_{2}=0
\\
M_{31}^{(n)}C_{1}+M_{32}^{(n)}C_{2}+M_{33}^{(n)}\tilde{C}_{1}+M_{34}^{(n)}\tilde{C}_{2}=0
\\
M_{41}^{(n)}C_{1}+M_{42}^{(n)}C_{2}+M_{43}^{(n)}\tilde{C}_{1}+M_{44}^{(n)}\tilde{C}_{2}=0\end{array}
\right\},\,n\geq1\eqno(45)
$$
and
$$
\left\{\begin{array}{l}[M_{11}^{(0)}\Theta(eB)+M_{22}^{(0)}\Theta(-eB)]C_{0}+[M_{13}^{(0)}\Theta(eB)+M_{24}^{(0)}\Theta(-eB)]\tilde{C}_{0}=0
\\
\,
[M_{31}^{(0)}\Theta(eB)+M_{42}^{(0)}\Theta(-eB)]C_{0}+[M_{33}^{(0)}\Theta(eB)+M_{44}^{(0)}\Theta(-eB)]\tilde{C}_{0}=0\end{array}
\right\},\eqno(46)
$$
where
$$
\left\{\begin{array}{l}M_{11}^{(n)}=\biggl[(\omega_{nl}+m)\left({\rm
e}^{{\vartheta}_{{\rm sgn}(eB)}}\cosh\tilde{\vartheta}_{{\rm
sgn}(eB)}\right)^{\Theta(E_{nl})} \\
+{\rm i}k_{l}\left({\rm
e}^{{\vartheta}_{{\rm sgn}(eB)}}\cosh\tilde{\vartheta}_{{\rm
sgn}(eB)}\right)^{\Theta(-E_{nl})}\biggr]{\rm e}^{{\rm i}k_{l}a/2}, \\
M_{12}^{(n)}={\rm i}\sqrt{2n|eB|}\left({\rm e}^{{\vartheta}_{{\rm
sgn}(eB)}}\cosh\tilde{\vartheta}_{{\rm
sgn}(eB)}\right)^{\Theta(-E_{nl})}{\rm
e}^{{\rm i}k_{l}a/2}, \\
M_{13}^{(n)}=M_{11}^{(n)*},M_{14}^{(n)}=-M_{12}^{(n)*}, \\
M_{21}^{(n)}=-{\rm i}\sqrt{2n|eB|}\left({\rm
e}^{-{\vartheta}_{{\rm sgn}(eB)}}\cosh\tilde{\vartheta}_{{\rm
sgn}(eB)}\right)^{\Theta(-E_{nl})}{\rm
e}^{{\rm i}k_{l}a/2}, \\
M_{22}^{(n)}=\biggl[(\omega_{nl}+m)\left({\rm
e}^{-{\vartheta}_{{\rm sgn}(eB)}}\cosh\tilde{\vartheta}_{{\rm
sgn}(eB)}\right)^{\Theta(E_{nl})} \\
+{\rm i}k_{l}\left({\rm e}^{-{\vartheta}_{{\rm
sgn}(eB)}}\cosh\tilde{\vartheta}_{{\rm
sgn}(eB)}\right)^{\Theta(-E_{nl})}\biggr]{\rm e}^{{\rm i}k_{l}a/2}, \\
M_{23}^{(n)}=-M_{21}^{(n)*},M_{24}^{(n)}=M_{22}^{(n)*}, \\
M_{31}^{(n)}=M_{33}^{(n)*},M_{32}^{(n)}=-M_{34}^{(n)*}, \\
M_{33}^{(n)}=\biggl[(\omega_{nl}+m)\left({\rm
e}^{-{\vartheta}_{-{\rm sgn}(eB)}}\cosh\tilde{\vartheta}_{-{\rm
sgn}(eB)}\right)^{\Theta(E_{nl})} \\
+{\rm i}k_{l}\left({\rm e}^{-{\vartheta}_{-{\rm
sgn}(eB)}}\cosh\tilde{\vartheta}_{-{\rm
sgn}(eB)}\right)^{\Theta(-E_{nl})}\biggr]{\rm e}^{{\rm i}k_{l}a/2}, \\
M_{34}^{(n)}=-{\rm i}\sqrt{2n|eB|}\left({\rm
e}^{-{\vartheta}_{-{\rm sgn}(eB)}}\cosh\tilde{\vartheta}_{-{\rm
sgn}(eB)}\right)^{\Theta(-E_{nl})}{\rm e}^{{\rm i}k_{l}a/2}, \\
M_{41}^{(n)}=-M_{43}^{(n)*},M_{42}^{(n)}=M_{44}^{(n)*}, \\
M_{43}^{(n)}={\rm i}\sqrt{2n|eB|}\left({\rm e}^{{\vartheta}_{-{\rm
sgn}(eB)}}\cosh\tilde{\vartheta}_{-{\rm
sgn}(eB)}\right)^{\Theta(-E_{nl})}{\rm e}^{{\rm
i}k_{l}a/2}, \\
M_{44}^{(n)}=\biggl[(\omega_{nl}+m)\left({\rm
e}^{{\vartheta}_{-{\rm sgn}(eB)}}\cosh\tilde{\vartheta}_{-{\rm
sgn}(eB)}\right)^{\Theta(E_{nl})} \\
+{\rm i}k_{l}\left({\rm e}^{{\vartheta}_{-{\rm
sgn}(eB)}}\cosh\tilde{\vartheta}_{-{\rm
sgn}(eB)}\right)^{\Theta(-E_{nl})}\biggr]{\rm e}^{{\rm i}k_{l}a/2}
\end{array}\right\}; \eqno(47)
$$
here the step function is defined as $\Theta(u)=1$ at $u>0$ and
$\Theta(u)=0$ at $u<0$, ${\rm sgn}(u)=\Theta(u)-\Theta(-u)$ is the
sign function, and we have introduced notations
$$
\omega_{nl}\equiv\omega_{nk_{l}}=\sqrt{2n|eB|+k_{l}^{2}+m^{2}}\eqno(48)
$$
and, similarly, $E_{nl} \equiv E_{nk_{l}}$.

Thus, the spectrum of wave number vector $k_{l}$ is determined
from condition
$$
\det M^{(n)}=0,\eqno(49)
$$
where
$$
\det M^{(n)}=(m+\omega_{nl})^{2}
$$
$$
\times\biggl\{{\rm e}^{2{\rm
i}k_{l}a}\biggl[m(\cosh^2\tilde{\vartheta}_{+} +1) + {\rm
sgn}(E_{nl})\omega_{nl}\sinh^2\tilde{\vartheta}_{+} + 2{\rm
i}k_{l}\cosh\tilde{\vartheta}_{+}\cosh\vartheta_{+}\biggr]
$$
$$
\times\biggl[m(\cosh^2\tilde{\vartheta}_{-} +1) + {\rm
sgn}(E_{nl})\omega_{nl}\sinh^2\tilde{\vartheta}_{-} + 2{\rm
i}k_{l}\cosh\tilde{\vartheta}_{-}\cosh\vartheta_{-}\biggr]
$$
$$
-2\biggl[m(\cosh^2\tilde{\vartheta}_{+} +1) + {\rm
sgn}(E_{nl})\omega_{nl}\sinh^2\tilde{\vartheta}_{+}\biggr]
$$
$$
\times\biggl[m(\cosh^2\tilde{\vartheta}_{-} +1) + {\rm
sgn}(E_{nl})\omega_{nl}\sinh^2\tilde{\vartheta}_{-}\biggr]
$$
$$
-4k_{l}^{2}(\cosh^2\tilde{\vartheta}_{+}+\cosh^2\tilde{\vartheta}_{-}+
2\cosh\tilde{\vartheta}_{+}\cosh\tilde{\vartheta}_{-}\sinh\vartheta_{+}\sinh\vartheta_{-})
$$
$$
+{\rm e}^{-2{\rm i}k_{l}a}\biggl[m(\cosh^2\tilde{\vartheta}_{+}
+1) + {\rm sgn}(E_{nl})\omega_{nl}\sinh^2\tilde{\vartheta}_{+} -
2{\rm i}k_{l}\cosh\tilde{\vartheta}_{+}\cosh\vartheta_{+}\biggr]
$$
$$
\times\biggl[m(\cosh^2\tilde{\vartheta}_{-} +1) + {\rm
sgn}(E_{nl})\omega_{nl}\sinh^2\tilde{\vartheta}_{-} - 2{\rm
i}k_{l}\cosh\tilde{\vartheta}_{-}\cosh\vartheta_{-}\biggr]\biggr\},n\geq1
\eqno(50)
$$
and
$$
\det
M^{(0)}\equiv[M_{11}^{(0)}\Theta(eB)+M_{22}^{(0)}\Theta(-eB)][M_{33}^{(0)}\Theta(eB)+M_{44}^{(0)}\Theta(-eB)]
$$
$$
-[M_{13}^{(0)}\Theta(eB)+M_{24}^{(0)}\Theta(-eB)][M_{31}^{(0)}\Theta(eB)+M_{42}^{(0)}\Theta(-eB)]
$$
$$
={\rm e}^{{\rm i}k_{l}a}\biggl[(m+\omega_{0l})\left({\rm
e}^{{\vartheta}_{+}}\cosh\tilde{\vartheta}_{+}\right)^{\Theta(E_{0l})}
 +{\rm i}k_{l}\left({\rm
e}^{{\vartheta}_{+}}\cosh\tilde{\vartheta}_{+}\right)^{\Theta(-E_{0l})}\biggr]
$$
$$
\times\biggl[(m+\omega_{0l})\left({\rm
e}^{-{\vartheta}_{-}}\cosh\tilde{\vartheta}_{-}\right)^{\Theta(E_{0l})}
 +{\rm i}k_{l}\left({\rm
e}^{-{\vartheta}_{-}}\cosh\tilde{\vartheta}_{-}\right)^{\Theta(-E_{0l})}\biggr]
$$
$$
- {\rm e}^{-{\rm i}k_{l}a}\biggl[(m+\omega_{0l})\left({\rm
e}^{{\vartheta}_{+}}\cosh\tilde{\vartheta}_{+}\right)^{\Theta(E_{0l})}
 -{\rm i}k_{l}\left({\rm
e}^{{\vartheta}_{+}}\cosh\tilde{\vartheta}_{+}\right)^{\Theta(-E_{0l})}\biggr]
$$
$$
\times\biggl[(m+\omega_{0l})\left({\rm
e}^{-{\vartheta}_{-}}\cosh\tilde{\vartheta}_{-}\right)^{\Theta(E_{0l})}
 -{\rm i}k_{l}\left({\rm
e}^{-{\vartheta}_{-}}\cosh\tilde{\vartheta}_{-}\right)^{\Theta(-E_{0l})}\biggr].\eqno(51)
$$

It should be recalled that, owing to boundary condition (44), the
normal component of current
$$
\boldsymbol{j}_{qnl}=\bar{\psi}_{qnl}(\mathbf{r})\boldsymbol{\gamma}\psi_{qnl}(\mathbf{r})\eqno
(52)
$$ vanishes at the boundary, see (21):
$$
j^{3}_{qnl}|_{z=\pm{a/2}}=0.\eqno(53)
$$
This signifies that the quantized matter is confined within the
boundaries.

Given solution $\psi_{q0l}^{(0)}(\mathbf{r})$, we impose the condition on its
coefficients $C_{0}$ and $\tilde{C}_{0}$:
$$
\left\{\begin{array}{l}|C_{0}|^{2}+|\tilde{C}_{0}|^{2}=\frac{2\pi}{a},\\
[3 mm] C_{0}^{*} \tilde{C}_{0}+\tilde{C}_{0}^{*}
C_{0}=0;\end{array}\right.\eqno(54)
$$
in particular, the coefficients can be chosen as
$$
C_{0}=\sqrt{\frac{\pi}{a}}{\rm e}^{\rm{i}\pi/4},\quad
\tilde{C}_{0}=\sqrt{\frac{\pi}{a}}{\rm e}^{-\rm{i}\pi/4}.\eqno(55)
$$
In the case of $n\geq1$, two linearly independent solutions,
$\psi_{qnl}^{(1)}(\mathbf{r})$ and $\psi_{qnl}^{(2)}(\mathbf{r})$,
are orthogonal, if the appropriate coefficients, $C_{j}^{(1)},\,
\tilde{C}_{j}^{(1)}$ and $C_{j}^{(2)},\, \tilde{C}_{j}^{(2)}$
$(j=1,2)$, obey condition
$$
\left\{\begin{array}{l}\sum\limits_{j=1,2}C_{j}^{(1)*}C_{j}^{(2)}=0,\\
[3 mm]
C_{j}^{(1)}C_{j'}^{(2)}=\tilde{C}_{j}^{(1)}\tilde{C}_{j'}^{(2)},\,
|C_{j}^{(j')}|=|\tilde{C}_{j}^{(j')}|,\quad
j,j'=1,2.\end{array}\right.\eqno(56)
$$
We impose further condition:
$$
\left\{\begin{array}{l}\sum\limits_{j=1,2}|C_{j}^{(j')}|^{2}=\frac{\pi}{a},\\
[6 mm] \sum\limits_{j=1,2} [C_{j}^{(j')*} \tilde{C}_{j}^{(j')} +
\tilde{C}_{j}^{(j')*}C_{j}^{(j')}]=0,\quad
j'=1,2;\end{array}\right.\eqno(57)
$$
in particular, the coefficients can be chosen as
$$
C_{1}^{(1)}=\sqrt{\frac{\pi}{2a}}{\rm
e}^{\rm{i}\pi/4},\,\tilde{C}_{1}^{(1)}=\sqrt{\frac{\pi}{2a}}{\rm
e}^{-\rm{i}\pi/4},\,C_{2}^{(1)}=\sqrt{\frac{\pi}{2a}}{\rm
e}^{-\rm{i}\pi/4},\,\tilde{C}_{2}^{(1)}=\sqrt{\frac{\pi}{2a}}{\rm
e}^{-3\rm{i}\pi/4}\eqno(58)
$$
and
$$
C_{1}^{(2)}=\sqrt{\frac{\pi}{2a}}{\rm
e}^{-\rm{i}\pi/4},\,\tilde{C}_{1}^{(2)}=\sqrt{\frac{\pi}{2a}}{\rm
e}^{\rm{i}\pi/4},\,C_{2}^{(2)}=\sqrt{\frac{\pi}{2a}}{\rm
e}^{\rm{i}\pi/4},\,\tilde{C}_{2}^{(2)}=\sqrt{\frac{\pi}{2a}}{\rm
e}^{3\rm{i}\pi/4}\eqno(59)
$$
As a result, wave functions
$\psi^{(j)}_{qnl}(\mathbf{r}) \, (j=0,1,2)$ satisfy the requirements
of orthonormality
$$\int\limits_{\Omega}{{\rm
d}^{3}r}\,{\psi^{(j)\dag}_{qnl}(\mathbf{r})}
\psi_{q'n'l'}^{(j')}(\mathbf{r})=\delta_{jj'}\delta_{nn'}\delta_{ll'}\delta(q-q'),
\quad j,j'=0,1,2 \eqno(60)
$$
and completeness
$$\sum\limits_{{\rm sgn}(E_{nl})}\,\int\limits_{-\infty}^{\infty}{\rm{d}}q\sum\limits_{l}\left[\psi^{(0)}_{q0l}(\mathbf{r})
\psi^{(0)\dag}_{q0l}(\mathbf{r'})+\sum\limits_{n=1}^{\infty}\sum\limits_{j=1,2}
\psi^{(j)}_{qnl}(\mathbf{r})\psi^{(j)\dag}_{qnl}(\mathbf{r'})\right]=I\delta(\mathbf{r}-\mathbf{r'}).\eqno(61)
$$
Consequently, we obtain the following formal expression for the
vacuum expectation value of the energy per unit area of the boundary
surface
$$
\frac{E}{S}\equiv\int\limits_{-a/2}^{a/2}{\rm{d}}z\,\varepsilon=
-\frac{|eB|}{2\pi}\sum\limits_{{\rm sgn}(E_{nl})}\sum\limits_{l}\sum\limits_{n=0}^{\infty}i_{n}\omega_{nl}.\eqno(62)
$$

\section{Determination of the spectrum of wave \\ number vector along the magnetic field}

The spectrum of wave number vector in the direction of the
magnetic field depends on four self-adjoint extension parameters,
$\vartheta_{+},\tilde{\vartheta}_{+},\vartheta_{-}$ and
$\tilde{\vartheta}_{-}$, see (44). In general, the values of these
self-adjoint extension parameters may vary arbitrarily from point
to point of the boundary surface. However, such a generality seems
to be excessive, lacking physical motivation, and we shall assume
in the following that the self-adjoint extension parameters are
independent of coordinates $x$ and $y$.

The equation determining the spectrum of $k_{l}$, see (49), can be
given in the form
$$
{\rm e}^{2{\rm i}k_{l}a}={\rm e}^{-2{\rm i}\eta_{k_{l}}},\eqno(63)
$$
or
$$
\sin(k_{l}a+\eta_{k_{l}})=0,\eqno(64)
$$
\newpage
where
$$
\eta_{k_l}=\frac 12{\rm
arctan}\frac{2k_{l}\cosh\tilde{\vartheta}_{+}\cosh\vartheta_{+}}{m(\cosh^2\tilde{\vartheta}_{+}
+1) + {\rm sgn}(E_{nl})\omega_{nl}\sinh^2\tilde{\vartheta}_{+}}
$$
$$
+\frac 12{\rm
arctan}\frac{2k_{l}\cosh\tilde{\vartheta}_{-}\cosh\vartheta_{-}}{m(\cosh^2\tilde{\vartheta}_{-}
+1) + {\rm sgn}(E_{nl})\omega_{nl}\sinh^2\tilde{\vartheta}_{-}}
\mp\frac 12{\rm arctan}\frac{2k_l\sqrt{\Delta}}{\beta},\quad n\geq
1, \eqno(65)
$$
$$
\Delta=\biggl\{\biggl[m(\cosh^2\tilde{\vartheta}_{+} +1) + {\rm
sgn}(E_{nl})\omega_{nl}\sinh^2\tilde{\vartheta}_{+}\biggr]\cosh\tilde{\vartheta}_{-}\sinh\vartheta_{-}
$$
$$
-\biggl[m(\cosh^2\tilde{\vartheta}_{-} +1) + {\rm
sgn}(E_{nl})\omega_{nl}\sinh^2\tilde{\vartheta}_{-}\biggr]\cosh\tilde{\vartheta}_{+}\sinh\vartheta_{+}\biggr\}^2
$$
$$
+4k_l^2\cosh\tilde{\vartheta}_{+}
\cosh\tilde{\vartheta}_{-}\biggl[\cosh\tilde{\vartheta}_{+}\cosh\tilde{\vartheta}_{-}(
\sinh\vartheta_{+}-\sinh\vartheta_{-})^2
$$
$$
-(\cosh\tilde{\vartheta}_{+}-\cosh\tilde{\vartheta}_{-})^2\sinh\vartheta_{+}\sinh\vartheta_{-}\biggr]
+2n|eB|(\sinh^2\tilde{\vartheta}_{+}-\sinh^2\tilde{\vartheta}_{-})^2,
\eqno(66)
$$
$$
\beta=\biggl[m(\cosh^2\tilde{\vartheta}_{+} +1) + {\rm
sgn}(E_{nl})\omega_{nl}\sinh^2\tilde{\vartheta}_{+}\biggr]
$$
$$
\times\biggl[m(\cosh^2\tilde{\vartheta}_{-} +1) + {\rm
sgn}(E_{nl})\omega_{nl}\sinh^2\tilde{\vartheta}_{-}\biggr]
$$
$$
+2k_{l}^{2}(\cosh^2\tilde{\vartheta}_{+}+\cosh^2\tilde{\vartheta}_{-}+
2\cosh\tilde{\vartheta}_{+}\cosh\tilde{\vartheta}_{-}\sinh\vartheta_{+}\sinh\vartheta_{-})
\eqno(67)
$$
[two signs in (65) correspond to two roots of the quadratic
equation for variable ${\rm e}^{2{\rm i}k_{l}a}$], and
$$
\eta_{k_l}=\arctan\biggl[\frac{k_{l}}{m+\omega_{0l}}\left({\rm
e}^{{\vartheta}_{+}}\cosh\tilde{\vartheta}_{+}\right)^{-{\rm
sgn}(E_{0l})}\biggr]
$$
$$
+\arctan\biggl[\frac{k_{l}}{m+\omega_{0l}}\left({\rm
e}^{-{\vartheta}_{-}}\cosh\tilde{\vartheta}_{-}\right)^{-{\rm
sgn}(E_{0l})}\biggr] \quad (n=0).\eqno(68)
$$

It should be noted that value $k_l=0$ is not permissible. Really,
we have in the case of $k_l=0$:
$$
\psi_{qnl}^{(j)}({\bf r})|_{z=a/2}=\psi_{qnl}^{(j)}({\bf
r})|_{z=-a/2}, \eqno(69)
$$
and boundary condition (44) can be presented in the form
$$
R \, \psi_{qnl}^{(j)}({\bf r})|_{k_l=0}=0, \eqno(70)
$$
where
$$ \left\{
\begin{array}{l}R_{11}=-{\rm
i}{\rm e}^{{\vartheta}_{+}}\cosh\tilde{\vartheta}_{+}, \quad
R_{12}=0, \quad R_{13}=1, \quad R_{14}=0, \\
R_{21}=0, \quad R_{22}={\rm i}{\rm
e}^{-{\vartheta}_{+}}\cosh\tilde{\vartheta}_{+}, \quad R_{23}=0, \quad R_{24}=1, \\
R_{31}={\rm i}{\rm
e}^{-{\vartheta}_{-}}\cosh\tilde{\vartheta}_{-}, \quad R_{32}=0,
\quad R_{33}=1, \quad R_{34}=0,
\\
R_{41}=0, \quad R_{42}=-{\rm i}{\rm
e}^{{\vartheta}_{-}}\cosh\tilde{\vartheta}_{-}, \quad R_{43}=0,
\quad R_{44}=1
\end{array}\right\}. \eqno(71)
$$
The determinant of matrix $R$ is nonzero at all values of the
self-adjoint extension parameters:
$$
\det R = \cosh^2\tilde{\vartheta}_{+} +
2\cosh\tilde{\vartheta}_{+}\cosh\tilde{\vartheta}_{-}\cosh({\vartheta}_{+}+{\vartheta}_{-})
+ \cosh^2\tilde{\vartheta}_{-}. \eqno(72)
$$
Hence, equation (70) allows for the trivial solution only,
$\psi_{qnl}^{(j)}({\bf r})|_{k_l=0}=0$; consequently, value
$k_l=0$ is excluded by the boundary condition.

It is not clear which of the signs in (65) should be chosen.
This ambiguity can be avoided by imposing restriction
$$
\vartheta_{+}=\vartheta_{-}=\vartheta, \quad
\tilde{\vartheta}_{+}=\tilde{\vartheta}_{-}=\tilde{\vartheta},\eqno(73)
$$
then (65) and (68) take form
$$
\eta_{k_{l}}=\arctan\frac{2k_{l}\cosh\tilde{\vartheta}\cosh\vartheta}{m(\cosh^2\tilde{\vartheta}
+1) + {\rm sgn}(E_{nl})\omega_{nl}\sinh^2\tilde{\vartheta}}, \quad
n \geq 0, \eqno(74)
$$
and the spectrum of $k_l$ consists of values of the
same sign, say, $k_l>0$; values of the opposite sign ($k_l<0$)
should be excluded to avoid double counting. Note that the
spectrum of $k_{l}$ depends on the number of the Landau level,
$n$, and on the sign of the one-particle energy, ${\rm sgn}(E_{nl})$, in this case as does in general.

By imposing further restriction
$$
\tilde{\vartheta}=0, \eqno(75)
$$
we arrive at the $k_{l}$-spectrum which is determined by condition
$$
\cos(k_{l}a)+\frac{m}{k_{l}\cosh\vartheta}\sin(k_{l}a)=0,
\eqno(76)
$$
being the same for all Landau levels and for both signs of the one-particle energy.
Note, that relations (20) and (22) in this case take forms
$$
(I \pm {\rm i}{\gamma}^3\cosh\vartheta + {\rm
i}\gamma^{5}\sinh\vartheta)\chi|_{z=\pm{a/2}}=0 \eqno(77)
$$
and
$$
\bar{\xi}(I\cosh\vartheta \mp
{\gamma}^3\gamma^{5}\sinh\vartheta)\chi|_{z=\pm{a/2}}=0, \eqno(78)
$$
respectively.

In the following our concern will be in the case of one
self-adjoint extension parameter with the $k_{l}$-spectrum that is
independent of $n$ and of  ${\rm sgn}(E_{nl})$, see (76).

\section{Casimir energy and force}

As was already mentioned, the expression for the induced vacuum
energy per unit area of the boundary surface, see (62), can be
regarded as purely formal, since it is ill-defined due to the
divergence of infinite sums over $l$ and $n$. To tame the
divergence, a factor containing a regularization parameter should
be inserted in (62). A summation over values $k_{l}>0$, which are
determined by (76), can be performed with the use of the
Abel-Plana formula and its generalizations \cite{Sah,Bel}. In the
case of $\cosh\vartheta=\infty$, which is formally equivalent to
the case of $m=0$, the well-known version of the Abel-Plana
formula (see, e.g., \cite{Bor}),
$$
\left.\sum\limits_{k_{l}>0}f(k_{l}^{2})\right|_{\cos(k_{l}a)=0}=
\frac{a}{2\pi}\int\limits_{-\infty}^{\infty}{\rm{d}}k{f(k^{2})}+\frac{{\rm
i}a}{\pi}\int\limits_{0}^{\infty}{\rm{d}}\kappa \frac{f[(-{\rm
i}\kappa)^{2})]-f[({\rm i}\kappa)^{2})]}{{\rm e}^{2{\kappa}a}+1},
\eqno(79)
$$
is used. In the case of $m/\cosh\vartheta>0$, the use is made of the following version of the Abel-Plana formula,
that is derived in Appendix B and which, at $\vartheta=0$, coincides after redefinition 
$f(\omega^{2}) \rightarrow f(\omega^{2})\left[1+\frac{m}{a(\omega^{2}+m^{2})}\right]$ 
with formula (15) in \cite{Bel},
$$
\sum\limits_{k_{l}>0}f(k_{l}^{2})=\frac{a}{2\pi}\int\limits_{-\infty}^{\infty}{\rm{d}}k{f(k^{2})}+\frac{{\rm
i}a}{\pi}\int\limits_{0}^{\infty}{\rm{d}}\kappa\Lambda(\kappa)\{f[(-{\rm
i}\kappa)^{2})]-f[({\rm i}\kappa)^{2})]\}
$$
$$
-\frac{1}{2}f(0)+\frac{m\cosh\vartheta}{2\pi}
\int\limits_{-\infty}^{\infty}{\rm{d}}k
\frac{f(k^{2})}{k^{2}\cosh^{2}\vartheta+m^{2}},\eqno(80)
$$
where
$$
\Lambda(\kappa)=\frac{\kappa\cosh\vartheta-m -
\frac{m\cosh\vartheta}{a(\kappa\cosh\vartheta+m)}
}{(\kappa\cosh\vartheta+m){\rm e}^{2\kappa
a}+\kappa\cosh\vartheta-m}.\eqno(81)
$$
Here, in (79) and (80), $f(\omega^{2})$ as a function of complex variable $\omega$
decreases sufficiently fast at large distances from the origin of
the complex $\omega$-plane. The regularization in the second term
on the right-hand side of (79) and (80) can be removed; then
$$
{\rm i}\{f[(-{\rm i}\kappa)^{2})]-f[({\rm
i}\kappa)^{2})]\}=-\frac{2|eB|}{\pi}\sum\limits_{n=0}^{\infty}i_{n}\sqrt{\kappa^{2}-2n|eB|-m^{2}}
\eqno(82)
$$
with the range of $\kappa$ restricted to
$\kappa>\sqrt{2n|eB|+m^{2}}$ for the corresponding terms. As to
the first term on the right-hand side of (79) and (80), one
immediately recognizes that it is equal to $\varepsilon^{\infty}$
(39) multiplied by $a$. Hence, if one ignores for a moment the
last terms on the right-hand side of (80), then the problem of
regularization and removal of the divergency in expression (62) is
the same as that in the case of no boundaries, when the magnetic
field fills the whole space. Defining the generalized Casimir
energy as the vacuum energy per unit area of the boundary surface,
which is renormalized in the same way as in the case of no
boundaries, we obtain at $m/\cosh\vartheta>0$:
$$
\frac{E_{\rm ren}}{S}=a\varepsilon^{\infty}_{\rm ren}-
\frac{2|eB|}{\pi^{2}}a\sum\limits_{n=0}^{\infty}i_{n}\int\limits_{M_{n}}^{\infty}{\rm
d}\kappa\Lambda(\kappa)\sqrt{\kappa^{2}-M_{n}^{2}}
$$
$$
+\frac{|eB|}{2\pi}\sum\limits_{n=0}^{\infty}i_{n}M_{n}-\frac{|eB|m\cosh\vartheta}{2\pi^{2}}\int\limits_{-\infty}^{\infty}{\rm
d}k
\sum\limits_{n=0}^{\infty}i_{n}\frac{\sqrt{k^{2}+M_{n}^{2}}}{k^{2}\cosh^{2}\vartheta+m^{2}},\eqno(83)
$$
where
$$
M_{n}=\sqrt{2n|eB|+m^{2}},\eqno (84)
$$
$\varepsilon^{\infty}_{\rm ren}$ is given by (40). The sums and
the integral in the last line on the right-hand side of (83)
[which are due to the terms in the last line on the right-hand
side of (80) and which can be interpreted as describing the proper
energies of the boundary planes containing the sources of the
magnetic field] are divergent, but this divergency is of no
concern for us, because it has no physical consequences. Rather
than the generalized Casimir energy, a physically relevant
quantity is the generalized Casimir force which is defined as
$$
F=-\frac{\partial}{\partial a}\frac{E_{\rm ren}}{S},\eqno(85)
$$
and which is free from divergencies. We obtain
$$
F=-\varepsilon^{\infty}_{\rm ren}-
\frac{2|eB|}{\pi^{2}}\sum\limits_{n=0}^{\infty}i_{n}\int\limits_{M_{n}}^{\infty}{\rm
d}\kappa\Upsilon(\kappa)\sqrt{\kappa^{2}-M_{n}^{2}},\eqno(86)
$$
where
$$
\Upsilon(\kappa)\equiv-\frac{\partial}{\partial a}a\Lambda(\kappa)
$$
$$
=\frac{\left[\left(2\kappa a-1\right)
\left(\kappa^{2}\cosh^{2}\vartheta-m^{2}\right)-2{\kappa}m\cosh\vartheta\right]{\rm
e}^{2{\kappa}a}-\left(\kappa\cosh\vartheta-m\right)^{2}}
{\left[\left(\kappa\cosh\vartheta+m\right){\rm
e}^{2{\kappa}a}+\kappa\cosh\vartheta-m\right]^{2}}.\eqno(87)
$$

It should be noted that limit $\cosh\vartheta \rightarrow \infty$
for $F$ (86) is smooth [i.e. the limiting value coincides with the
result obtained with the use of (79)]. Thus, for a particular
choice of the boundary condition yielding spectrum
 $k_{l}=\frac{\pi}{a}(l+\frac{1}{2})$ $(l=0,1,2,...)$,
we obtain
$$
\left.\frac{E_{\rm
ren}}{S}\right|_{\vartheta=\pm\infty}=a\varepsilon^{\infty}_{\rm
ren}-
\frac{2|eB|}{\pi^{2}}a\sum\limits_{n=0}^{\infty}i_{n}\int\limits_{M_{n}}^{\infty}{\rm
d}\kappa\frac{\sqrt{\kappa^{2}-M_{n}^{2}}}{{\rm
e}^{2{\kappa}a}+1}\eqno(88)
$$
and, using integration by parts,
$$
\left.F\right|_{\vartheta=\pm\infty}=-\varepsilon^{\infty}_{\rm
ren}-
\frac{2|eB|}{\pi^{2}}\sum\limits_{n=0}^{\infty}i_{n}\int\limits_{M_{n}}^{\infty}\frac{{\rm
d}\kappa}{{\rm
e}^{2{\kappa}a}+1}\frac{\kappa^{2}}{\sqrt{\kappa^{2}-M_{n}^{2}}}.\eqno(89)
$$
The integral in (88) can be taken after expanding the factor with
denominator as $\sum\limits_{j=1}^{\infty}(-1)^{j-1}{\rm
e}^{-2j{\kappa}a}$ . In this way, we obtain the following
expressions for the Casimir energy
$$
\left.\frac{E_{\rm
ren}}{S}\right|_{\vartheta=\pm\infty}=a\varepsilon^{\infty}_{\rm
ren}-
\frac{|eB|}{\pi^{2}}\sum\limits_{n=0}^{\infty}i_{n}M_{n}\sum\limits_{j=1}^{\infty}(-1)^{j-1}\frac{1}{j}
K_{1}(2jM_{n}a)\eqno(90)
$$
and the Casimir force
$$
\left.F\right|_{\vartheta=\pm\infty}=-\varepsilon^{\infty}_{\rm
ren}-
\frac{2|eB|}{\pi^{2}}\sum\limits_{n=0}^{\infty}i_{n}M_{n}^{2}\sum\limits_{j=1}^{\infty}(-1)^{j-1}
\left[K_{0}(2jM_{n}a)+\frac{1}{2jM_{n}a}K_{1}(2jM_{n}a)\right],\eqno(91)
$$
where $K_{\rho}(u)$ is the Macdonald function of order $\rho$. The
case of $\vartheta=\pm\infty$, as was already mentioned, is
formally equivalent to the case of a massless spinor field, $m=0$;
however, it has to be kept in mind that the
$\vartheta$-independent piece of the Casimir force,
$-\varepsilon^{\infty}_{\rm ren}$, diverges in the limit of $m
\rightarrow 0$, see (40).

Note also that the antiperiodic boundary condition,
$$
\chi|_{z=a/2}+\chi|_{z=-a/2}=0\eqno(92)
$$
(the same condition is for $\xi$), ensures the self-adjointness of
the Dirac Hamiltonian operator, but current (52) does not vanish
at the boundary: instead, the influx of the quantized matter at
one boundary surface equals the outflux of the quantized matter at
the other boundary surface. The spectrum of the wave number vector
which is orthogonal to the boundary is
$k_{l}=\frac{2\pi}{a}(l+\frac{1}{2})$ $(l=0,\pm1,\pm2,...)$, and
the Casimir energy and force take forms
$$
\left(\frac{E_{\rm ren}}{S}\right)_{\rm
antiperiod}=a\varepsilon^{\infty}_{\rm ren}-
\frac{2|eB|}{\pi^{2}}a\sum\limits_{n=0}^{\infty}i_{n}\int\limits_{M_{n}}^{\infty}{\rm
d}\kappa\frac{\sqrt{\kappa^{2}-M_{n}^{2}}}{{\rm
e}^{{\kappa}a}+1}\eqno(93)
$$
and
$$
(F)_{\rm antiperiod}=-\varepsilon^{\infty}_{\rm ren}-
\frac{2|eB|}{\pi^{2}}\sum\limits_{n=0}^{\infty}i_{n}\int\limits_{M_{n}}^{\infty}\frac{{\rm
d}\kappa}{{\rm
e}^{{\kappa}a}+1}\frac{\kappa^{2}}{\sqrt{\kappa^{2}-M_{n}^{2}}},\eqno(94)
$$
respectively, or, in the alternative representation,
$$
\left(\frac{E_{\rm ren}}{S}\right)_{\rm
antiperiod}=a\varepsilon^{\infty}_{\rm ren}-
\frac{2|eB|}{\pi^{2}}\sum\limits_{n=0}^{\infty}i_{n}M_{n}\sum\limits_{j=1}^{\infty}(-1)^{j-1}\frac{1}{j}
K_{1}(jM_{n}a),\eqno(95)
$$
and
$$
(F)_{\rm antiperiod}=-\varepsilon^{\infty}_{\rm ren}-
\frac{2|eB|}{\pi^{2}}\sum\limits_{n=0}^{\infty}i_{n}M_{n}^{2}\sum\limits_{j=1}^{\infty}(-1)^{j-1}
\left[K_{0}(jM_{n}a)+\frac{1}{jM_{n}a}K_{1}(jM_{n}a)\right].\eqno(96)
$$

\section{Conclusion and discussion}

  In the present paper, we have considered the influence of a
background uniform magnetic field and boundary conditions on the
vacuum of a quantized charged spinor matter field confined between
two parallel plates separated by distance $a$. If the magnetic
field is directed orthogonally to the plates and the normal
component of the current of quantized matter is assumed to vanish
at the plates, then the Dirac Hamiltonian operator is self-adjoint
under a set of boundary conditions depending on four arbitrary
functions of two coordinates which are tangential to the plates.
Ignoring this functional dependence and restricting ourselves to
the case when the spectrum of the wave number vector along the
magnetic field is independent of the number of the Landau level,
see (76), we arrive at a set of boundary conditions depending on
one parameter, $\vartheta$, see (77). Under these circumstances
the Casimir force is shown to take the form of (86), where
$\varepsilon^{\infty}_{\rm {ren}}$ is given by (40) and
$\Upsilon(\kappa)$ is given by (87). For a particular boundary
condition, $\vartheta=\pm\infty$, the Casimir force is given by
(89) or, alternatively, by (91). The latter is to be compared with
the case of the antiperiodic boundary condition, see (92), when
the normal component of the current is not vanishing at the
boundary and the Casimir force takes the form of (94) or,
alternatively, of (96).

In the limit of a weak magnetic field, $|B|\ll{m^{2}|e|^{-1}}$,
one has (see \cite{Hei2})
$$
\varepsilon^{\infty}_{\rm
{ren}}=-\frac{1}{360\pi^{2}}\frac{e^{4}B^{4}}{m^{4}}.\eqno(97)
$$
Thus, at $|B|\rightarrow0$ the first term on the right-hand side
of (86) vanishes, and, substituting the sum in the remaining piece
there by integral $\int\limits_{0}^{\infty}{\rm d}n$, we get
$$
F+\varepsilon^{\infty}_{\rm ren}=-
\frac{2}{3\pi^{2}}\int\limits_{m}^{\infty}{\rm
d}\kappa\Upsilon(\kappa)(\kappa^{2}-m^{2})^{3/2},\quad
|eB|\ll{m^{2}},\eqno(98)
$$
which in the limits of large and small distances between the plates
take the forms:
$$
F+\varepsilon^{\infty}_{\rm ren}=
\left\{\begin{array}{l}-\frac{3}{16\pi^{3/2}}\frac{m^{3/2}}{a^{5/2}}{\rm
e}^{-2ma}[1+O(\frac{1}{ma})],\quad \vartheta=0
\\ [6 mm]
-\frac{\tanh^{2}(\vartheta/2)}{2\pi^{3/2}}\frac{m^{5/2}}{a^{3/2}}{\rm
e}^{-2ma}[1+O(\frac{1}{ma})],\quad \vartheta \neq 0
\end{array}
\right\},\, ma\gg1\eqno(99)
$$
and
$$
F+\varepsilon^{\infty}_{\rm
ren}=-\frac{7}{8}\frac{\pi^{2}}{120}\frac{1}{a^{4}},\quad
ma\ll1. \eqno(100)
$$
Result (99) at $\vartheta=0$ is already known, see \cite{Most}, as well as result (100) is for a long time \cite{Joh} (the latter equals $1/16$ of
the appropriate result in the case of the antiperiodic boundary condition \cite{For}).

In the limit of a strong magnetic field, $|B|\gg{m}^{2}|e|^{-1}$, one has (see, e.g., \cite{Dun})
$$
\varepsilon^{\infty}_{\rm
ren}=-\frac{e^{2}B^{2}}{24\pi^{2}}\ln\frac{2|eB|}{m^{2}},\eqno(101)
$$
while the remaining piece of the force is
$$
F+\varepsilon^{\infty}_{\rm ren}=-
\frac{|eB|}{\pi^{2}}\left[\int\limits_{m}^{\infty}{\rm
d}\kappa\Upsilon(\kappa)\sqrt{\kappa^{2}-m^{2}}
+2\sum\limits_{n=1}^{\infty}\int\limits_{\sqrt{2n|eB|}}^{\infty}\frac{{\rm
d}\kappa}{{\rm
e}^{2{\kappa}a}+1}\frac{\kappa^{2}}{\sqrt{\kappa^{2}-2n|eB|}}\right].\eqno(102)
$$
The latter expression in the limit of large distances between the plates take forms
$$
F+\varepsilon^{\infty}_{\rm ren}=
\left\{\begin{array}{l}-\frac{|eB|}{16\pi^{3/2}}\frac{m^{1/2}}{a^{3/2}}{\rm
e}^{-2ma}[1+O(\frac{1}{ma})],\quad \vartheta=0
\\ [6 mm]
-\frac{|eB|\tanh^{2}(\vartheta/2)}{2\pi^{3/2}}\frac{m^{3/2}}{a^{1/2}}{\rm
e}^{-2ma}[1+O(\frac{1}{ma})],\quad \vartheta \neq 0 \end{array}
\right\},
$$
$$
\sqrt{|eB|}a\gg{ma}\gg1\eqno(103)
$$
and
$$
F+\varepsilon^{\infty}_{\rm ren}=-
\frac{|eB|}{\pi^{2}}\int\limits_{m}^{\infty}{\rm
d}\kappa\Upsilon(\kappa)\sqrt{\kappa^{2}-m^{2}}-\left(\frac{\sqrt{2}}{\pi}\right)^{3/2}\frac{|eB|^{7/4}}{\sqrt{a}}{\rm
e}^{-2\sqrt{2|eB|}a}$$
$$\times\biggl\{1+O[(\sqrt{|eB|}a)^{-1}]
+O[{\rm e}^{-2(\sqrt{2}-1)\sqrt{2|eB|}a}]\biggr\},\quad
\sqrt{|eB|}a\gg1,\eqno(104)
$$
while, in the limit of small distances between the plates, we get
$$
F+\varepsilon^{\infty}_{\rm ren}=- \frac{|eB|}{48a^{2}},\quad
ma\ll1,\,\sqrt{|eB|}a\gg1\eqno(105)
$$
and (100) at $ma\ll\sqrt{|eB|}a\ll1$. Result (103) at $\vartheta=0$ and result (105) were
obtained erarlier in \cite{Eli}; the latter of the results equals 1/4 of
the appropriate result in the case of the antiperiodic boundary condition \cite{Cou}.

We can conclude that the Heisenberg-Euler term, $\varepsilon_{\rm
ren}^\infty$ (40), is dominating at a relatively large separation
of the plates, $a\gg m^{-1}$, at a nonweak magnetic field. Since
the right-hand side of (40) is negative, the Casimir force in this
case, $F\approx -\varepsilon_{\rm ren}^\infty$, is repulsive (the
pressure from the vacuum onto the plates is positive), being
independent of the choice of boundary conditions at the plates, as
well as of the distance between the plates. In the opposite case
of a relatively small separation of the plates, $a\ll m^{-1}$, at
a sufficiently weak magnetic field, $|B| \ll m^2|e|^{-1}$, the
Heisenberg-Euler term is negligible, and the Casimir force is
attractive, being power dependent on the distance between the
plates, see (100) and (105). We remind that the results for the
case of the MIT bag boundary condition are obtained at
$\vartheta=0$, while the results for the case of the antiperiodic
boundary condition are obtained at $\vartheta=\pm\infty$ by change
$a\rightarrow{a/2}$.

Let us compare our results with those of the authors of
\cite{Cou,Eli,Ost}. As was already mentioned in Introduction,
these authors assume that both the quantized and external fields
are not confined within the plates. Due to this circumstance, the
Heisenberg-Euler-term contribution to the Casimir effect is absent
in their approach. As to the remaining part, it is calculated in
the cases of the antiperiodic boundary condition \cite{Cou,Ost}
and the MIT bag boundary condition \cite{Eli}; our results for
$F+\varepsilon^{\infty}_{\rm ren}$ in these particular cases agree
with the results in \cite{Cou,Eli,Ost}. Note also that the Casimir
effect with a quantized charged scalar matter field in the
background of an external uniform magnetic field has been
comprehensively analyzed in \cite{Sit}.

Usually, the Casimir effect is validated experimentally for the
separation of parallel plates to be of order of $10^{-8}-10^{-5}\,
\rm m$, see, e.g., Ref.\cite{Bor}.  So, even if one takes the
lightest charged particle, electron (the Compton wavelength,
$\lambda_{C}=\hbar(mc)^{-1}$, equals $3.86\times10^{-13}\,\rm m$),
then it becomes clear that the limiting case of
$a\lambda^{-1}_{C}\ll1$ [which is appropriate to (100) and (105),
when constants $\hbar$ and $c$ are recovered] has no relation to
physics reality. In the realistic case of $a\lambda^{-1}_{C}\gg1$
the Casimir force is prevailed by the Heisenberg-Euler term,
$F\approx-\varepsilon_{\rm ren}^\infty$ , since the corrections
depending on the separation distance and boundary conditions are
exponentially damped, see (99) and (103). Thus, in the limit of a
strong magnetic field, $|B|{\gg}B_{\rm crit}$, we obtain
$$
F=\frac{1}{24\pi^{2}}\frac{{\hbar}c}{\lambda_{C}^{4}}\left(\frac{B}{B_{\rm
crit}}\right)^{2}\ln\frac{2|B|}{B_{\rm crit}},\eqno (106)
$$
where constants $\hbar$ and $c$ are recovered, and $B_{\rm
crit}={\hbar}c(\lambda_{C}^{2}|e|)^{-1}$ equals
$4.41\times10^{13}\,\rm G$. Such supercritical magnetic fields may
be attainable in some astrophysical objects, such as neutron stars
and magnetars \cite{Har}, and also gamma-ray bursters in scenarios
involving protomagnetars \cite{Met}; the proper account for the
influence of Casimir pressure (106) on physical processes in these
objects should be taken.

Supercritical magnetic fields are not feasible in terrestrial
laboratories where the maximal values of steady magnetic fields
are of order of $10^{5}\,\rm G$, see, e.g., \cite{Per}. In the
case of a subcritical magnetic field, $|B|{\ll}B_{\rm crit}$, we
obtain by rewriting (97):
$$
F=\frac{1}{360\pi^{2}}\frac{{\hbar}c}{\lambda_{C}^{4}}\left(\frac{B}{B_{\rm
crit}}\right)^{4}. \eqno (107)
$$
Let us compare this with the attractive Casimir force which is due
to the quantized electromagnetic field \cite{Cas1},
$$
 F^{(EM)}=-\frac{\pi^{2}}{240}\frac{{\hbar}c}{a^{4}}, \eqno(108)
$$
and define ratio
$$
\frac{F}{F^{(EM)}}=-\frac{2}{3\pi^{4}}\left(\frac{a}{\lambda_{C}}\right)^{4}\left(\frac{B}{B_{\rm
crit}}\right)^{4}. \eqno (109)
$$
At $a=10^{-6}\,\rm m$ and $B=10^{5}\,\rm G$ the attraction is
prevailing over the repulsion by six orders of magnitude,
$F^{(EM)}/F\approx-10^{6}$, and the Casimir force is
$F^{(EM)}\approx-1.3\,\rm mPa$. However, at $a=10^{-4}\,\rm m$ and
$B=10^{5}\,\rm G$ the repulsion becomes dominant over the
attraction by two orders of magnitude, $F/F^{(EM)}\approx-10^{2}$
and the Casimir force takes value $F \approx 0.009\,\rm mPa$.
Otherwise, the same value of the Casimir force is achieved at
$a=10^{-5}\,\rm m$ and $B=10^{6}\,\rm G$. Thus, an experimental
observation of the influence of the background magnetic field on
the Casimir pressure seems to be possible in a foreseen future in
terrestrial laboratories.

\section*{Acknowledgments}

I am thankful to F.~Niedermayer and U.-J.~Wiese for interesting
discussions, and to S.A.~Yushchenko for collaboration on initial
stages of the work. The research was supported by the National
Academy of Sciences of Ukraine (project No. 0112U000054) and by the
State Fund for Fundamental Researches of Ukraine (SFFR-BRFFR Grant
No. F54.1/019). A partial support from the Program of Fundamental
Research of the Department of Physics and Astronomy of the
National Academy of Sciences of Ukraine (project No. 0112U000056)
and from the ICTP -- SEENET-MTP Grant No. PRJ-09 ``Strings and
Cosmology'' is also acknowledged.

\section*{Appendix A. Solution to the Dirac equation with negative energy}

The solution with negative energy, $E_{nk}=-\omega_{nk}$,
takes the following form in the case of $eB>0$ (magnetic field is directed along the $z$-axis):
$$\psi_{qnk}(\mathbf{r})=\frac{{\rm e}^{-{\rm i}qx}{\rm e}^{-{\rm i}kz}}{2\pi\sqrt{2\omega_{nk}(\omega_{nk}+m)}}
\left[\tilde{C}_{1}\begin{pmatrix} kY_{n}^{-q}(y)
\\-\sqrt{2neB}Y_{n-1}^{-q}(y) \\ (\omega_{nk}+m)Y_{n}^{-q}(y) \\ 0
\end{pmatrix}\right.$$
$$\left. + \tilde{C}_{2}\begin{pmatrix}-\sqrt{2neB}Y_{n}^{-q}(y)\\
-kY_{n-1}^{-q}(y) \\ 0 \\(\omega_{nk}+m)Y_{n-1}^{-q}(y)
\end{pmatrix}\right],\quad n\geq1\eqno(A.1)
$$
and
$$\psi^{(0)}_{q0k}(\mathbf{r})=\frac{{\rm e}^{-{\rm i}qx}{\rm e}^{-{\rm i}kz}}{2\pi\sqrt{2\omega_{0k}(\omega_{0k}+m)}}\tilde{C}_{0}Y_{0}^{-q}(y)
\begin{pmatrix} k \\ 0 \\ \omega_{0k}+m \\ 0 \end{pmatrix}. \eqno(A.2)
$$
The solution
corresponding to the plane wave propagating along the $z$-axis in
the opposite direction is obtained from (A.1) and (A.2) by
changing $k\rightarrow-k$ (coefficients
$\tilde{C}_{0},\tilde{C}_{1},\tilde{C}_{2}$ should be changed to
$C_{0},C_{1},C_{2}$).

\section*{Appendix B. Abel-Plana summation formula}

Let us rewrite condition (76) as
$$
P(k_{l})=0,\eqno(B.1)
$$
where
$$
P(k)=\cos(ka)+\frac{m}{k\cosh\vartheta}\sin(ka).\eqno(B.2)
$$
We assign labels $l=0,1,2,...$, to the consecutively increasing
positive roots of (B.1), $k_{l}>0$; appropriately, labels
$l=-1,-2,...$, are assigned to the consecutively decreasing
negative roots of (B.1), $k_{l}<0$. Then one can write
$$\sum\limits_{l=0}^{\infty}f(k_{l}^{2})=\frac{1}{2}\sum\limits_{l=-\infty}^{\infty}f(k_{l}^{2})=\frac{a}{4\pi}\int\limits_{C_{\underline{\overline{\,\,\,\,}}}}{\rm d}\omega{f(\omega^{2})}G(\omega),\eqno(B.3)
$$
where
$$
G(\omega)=1+\frac{\rm i}{a}\frac{\rm d}{{\rm
d}\omega}\ln{P(\omega)}\eqno(B.4)
$$
\begin{figure}
\includegraphics[width=390pt]{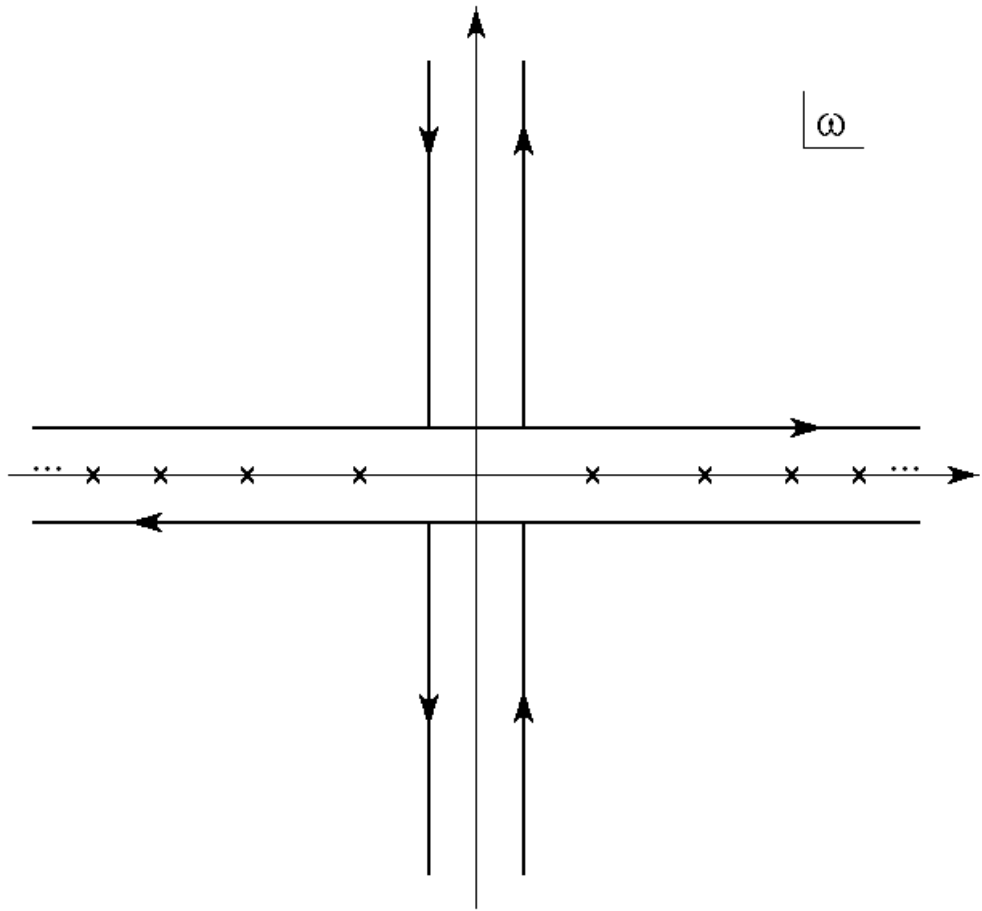}\\
\caption{Contours $C_{\underline{\overline{\,\,\,\,}}}$, $C_{\sqcap}$ and $C_{\sqcup}$ on
the complex $\omega$-plane; the positions of poles of $G(\omega)$ are
indicated by crosses.}\label{1}
\end{figure}\normalsize
and contour $C_{\underline{\overline{\,\,\,\,}}}$ on the complex
$\omega$-plane consists of two parallel infinite lines going closely
on the lower and upper sides of the real axis, see Fig.1. By
deforming the parts of contour $C_{\underline{\overline{\,\,\,\,}}}$ into contours $C_{\sqcap}$
and $C_{\sqcup}$ enclosing the lower and upper imaginary semiaxes,
see Fig.1, we get
$$
\int\limits_{C_{\underline{\overline{\,\,\,\,}}}}{\rm
d}\omega{f(\omega^{2})}G(\omega)=\int\limits_{C_{\sqcap}}{\rm
d}\omega{f(\omega^{2})}G(\omega)+\int\limits_{C_{\sqcup}}{\rm
d}\omega{f(\omega^{2})}G(\omega),\eqno(B.5)
$$
where it is implied that all singularities of $f$ as a function of
complex variable $\omega$ lie on the imaginary axis at some
distances from the origin. In view of obvious relation
$$\lim\limits_{k\rightarrow{0+}}(k\pm{{\rm i}\kappa})^{2}=\lim\limits_{k\rightarrow{0+}}(-k\mp{{\rm i}\kappa})^{2}=(\pm{{\rm i}\kappa})^{2}
$$
for real positive $k$ and $\kappa$, the right-hand side of (B.5) is
rewritten in the following way:
$$
\int\limits_{C_{\underline{\overline{\,\,\,\,}}}}{\rm d}\omega{f(\omega^{2})}G(\omega)={\rm
i}\int\limits_{0}^{\infty}{\rm{d}}\kappa\{f[(-{\rm
i}\kappa)^{2}]-f[({\rm i}\kappa)^{2}]\}[G(-{\rm i}\kappa)-G({\rm
i}\kappa)].\eqno(B.6)
$$
Taking account for relation
$$
G(-{\rm i}\kappa)+G({\rm i}\kappa)=2,\eqno(B.7)
$$
\newpage  we further obtain
$$
\int\limits_{C_{\underline{\overline{\,\,\,\,}}}}{\rm d}\omega{f(\omega^{2})}G(\omega)=2{\rm
i}\int\limits_{0}^{\infty}{\rm{d}}\kappa\{f[(-{\rm
i}\kappa)^{2})]-f[({\rm i}\kappa)^{2})]\}G(-{\rm i}\kappa)
$$
$$
-2{\rm i}\int\limits_{0}^{\infty}{\rm{d}}{\kappa}f[(-{\rm
i}\kappa)^{2})]+2{\rm
i}\int\limits_{0}^{\infty}{\rm{d}}{\kappa}f[({\rm
i}\kappa)^{2})].\eqno(B.8)
$$
By rotating the paths of integration in the last and before the last
integrals in (B.8) by $90^{\circ}$ in the clockwise and
anticlockwise directions, respectively, we finally get
$$
\int\limits_{C_{\underline{\overline{\,\,\,\,}}}}{\rm d}\omega{f(\omega^{2})}G(\omega)=2{\rm
i}\int\limits_{0}^{\infty}{\rm{d}}\kappa\{f[(-{\rm
i}\kappa)^{2})]-f[({\rm i}\kappa)^{2})]\}G(-{\rm
i}\kappa)+4\int\limits_{0}^{\infty}{\rm{d}}k f(k^{2}).\eqno(B.9)
$$

Note that the explicit form of function $G(\omega)$ is
$$
G(\omega)=\frac{(\omega\cosh\vartheta+{\rm i}m){\rm e}^{-{\rm
i}{\omega}a} - {\rm i}m
\frac{\sin({\omega}a)}{{\omega}a}}{{\omega}\cosh\vartheta\cos({\omega}a)+m
\sin({\omega}a)}.\eqno(B.10)
$$
Since the numerator of $G(\omega)$ (B.10) contributes to the
integral on the left-hand side of (B.9) at values $\omega=k_{l}$
only, one may change the numerator with the use of relation (B.1).
We can employ this arbitrariness and change $G(\omega)$ to $\tilde{G}(\omega)$ in such a
way that $\tilde{G}(-{\rm i}\kappa)$ will become exponentially decreasing
at large values of $\kappa$. Namely, we make substitution
$$
m\frac{\sin({\omega}a)}{{\omega}a}\rightarrow-\frac{m^{2}}{{\omega}^{2}\cosh^{2}\vartheta+m^{2}}
\,\frac{\cosh\vartheta}{a\cos({\omega}a)}. \eqno(B.11)
$$
However, then additional simple poles appear at $\omega=0$ and
at $\cos({\omega}a)=0$. Subtracting the contribution of these poles,
we obtain
$$
\int\limits_{C_{\underline{\overline{\,\,\,\,}}}}{\rm
d}\omega{f(\omega^{2})}G(\omega)=2{\rm
i}\int\limits_{0}^{\infty}{\rm{d}}\kappa\{f[(-{\rm
i}\kappa)^{2})]-f[({\rm i}\kappa)^{2})]\}\tilde{G}(-{\rm
i}\kappa)+4\int\limits_{0}^{\infty}{\rm{d}}k f(k^{2})
$$
$$
-\frac{\rm
i}{a}\int\limits_{C_{\underline{\overline{\,\,\,\,}}}}{\rm
d}\omega\frac{f(\omega^{2})}{\omega} +
\frac{m\cosh\vartheta}{a}\int\limits_{C_{\underline{\overline{\,\,\,\,}}}}{\rm
d}\omega\frac{f(\omega^{2})}{{\omega}^{2}\cosh^{2}\vartheta+m^{2}}
\frac{{\rm e}^{-{\rm i}{\omega}a}}{\cos({\omega}a)},\eqno(B.12)
$$
where
$$
\tilde{G}(\omega)=\frac{(\omega\cosh\vartheta+{\rm i}m){\rm
e}^{-{\rm i}{\omega}a} + \frac{{\rm
i}m^{2}}{{\omega}^{2}\cosh^{2}\vartheta+m^{2}}
\,\frac{\cosh\vartheta}{a\cos({\omega}a)}}{{\omega}\cosh\vartheta\cos({\omega}a)+m
\sin({\omega}a)}.\eqno(B.13)
$$
The last integral on the right-hand side of (B.12) is transformed
into integrals along the imaginary axis on the complex
$\omega$-plane in the same manner as previously, see (B.5)-(B.9).
In this way we get
$$
\int\limits_{C_{\underline{\overline{\,\,\,\,}}}}{\rm
d}\omega{f(\omega^{2})}G(\omega)=4{\rm
i}\int\limits_{0}^{\infty}{\rm{d}}\kappa\{f[(-{\rm
i}\kappa)^{2})]-f[({\rm
i}\kappa)^{2})]\}\Lambda(\kappa)+4\int\limits_{0}^{\infty}{\rm{d}}k
f(k^{2})
$$
$$
-\frac{2\pi}{a}f(0) +
4\frac{m\cosh\vartheta}{a}\int\limits_{0}^{\infty}{\rm{d}}k
\frac{f(k^{2})}{k^{2}\cosh^{2}\vartheta+m^{2}}, \eqno(B.14)
$$
where the contribution of the pole at $\omega=0$ is explicitly
written, and
$$
\Lambda(\kappa)=\frac{1}{2}\biggl[\tilde{G}(-{\rm i}\kappa) -
\frac{1}{a}\,\,\frac{m\cosh\vartheta}{\kappa^{2}\cosh^{2}\vartheta-m^{2}}\,\,\frac{{\rm
e}^{-{\kappa}a}}{\cosh({\kappa}a)}\biggr] \eqno(B.15)
$$
is explicitly given by (81). It should be noted that the
contribution of poles on the imaginary axis at $\omega=\pm {\rm i}
m/\cosh\vartheta$, stemming from substitution (B.11), is
canceled. Recalling (B.3), we rewrite (B.14) into the form given by (80).

\end{document}